\documentclass[acmtog]{acmart}
\acmSubmissionID{1381}

\usepackage{bbold}
\usepackage{algorithm2e}

\acmJournal{TOG}

\copyrightyear{2025}
\acmYear{2025}
\setcopyright{cc}
\setcctype{by}
\acmConference[SIGGRAPH Conference Papers '25]{Special Interest Group on Computer Graphics and
Interactive Techniques Conference Conference Papers }{August 10--14, 2025}{Vancouver, BC, Canada}
\acmBooktitle{Special Interest Group on Computer Graphics and Interactive Techniques Conference
Conference Papers (SIGGRAPH Conference Papers '25), August 10--14, 2025, Vancouver, BC, Canada}
\acmDOI{10.1145/3721238.3730756}
\acmISBN{979-8-4007-1540-2/2025/08}

\begin{document}
\title{ELGAR: Expressive Cello Performance Motion Generation for Audio Rendition}

\author{Zhiping Qiu}
\orcid{0009-0006-8663-1955}
\affiliation{%
 \institution{Central Conservatory of Music}
 \country{China}}
\affiliation{%
 \institution{Tsinghua University}
 \country{China}}
\email{zhiping_qiu@mail.ccom.edu.cn}

\author{Yitong Jin}
\orcid{0009-0002-3979-5878}
\affiliation{%
 \institution{Central Conservatory of Music}
 \country{China}}
\affiliation{%
 \institution{Tsinghua University}
 \country{China}}
\email{jinyitong@mail.ccom.edu.cn}

\author{Yuan Wang}
\orcid{0009-0007-8604-6243}
\affiliation{%
 \institution{Central Conservatory of Music}
 \country{China}}
\email{22a056@mail.ccom.edu.cn}

\author{Yi Shi}
\orcid{0000-0001-7500-192X}
\affiliation{%
 \institution{Central Conservatory of Music}
 \country{China}}
\affiliation{%
 \institution{Tsinghua University}
 \country{China}}
\email{shiyi@mail.ccom.edu.cn}

\author{Chongwu Wang}
\orcid{0009-0002-1384-941X}
\affiliation{%
 \institution{Central Conservatory of Music}
 \country{China}}
\email{1225@ccom.edu.cn}

\author{Chao Tan}
\orcid{0009-0006-2472-0258}
\affiliation{%
 \institution{Weilan Tech}
 \country{China}}
\email{chao.tan333@139.com}

\author{Xiaobing Li}
\orcid{0000-0003-0113-824X}
\affiliation{%
 \institution{Central Conservatory of Music}
 \country{China}}
\email{lxiaobing@ccom.edu.cn}

\author{Feng Yu}
\orcid{0009-0007-0607-7315}
\affiliation{%
 \institution{Central Conservatory of Music}
 \country{China}}
\email{yufengAI@ccom.edu.cn}

\author{Tao Yu}
\orcid{0000-0002-3818-5069}
\affiliation{%
 \institution{Tsinghua University}
 \country{China}}
\email{ytrock@mail.tsinghua.edu.cn}
\authornote{Corresponding Author}

\author{Qionghai Dai}
\orcid{0000-0001-7043-3061}
\affiliation{%
 \institution{Tsinghua University}
 \country{China}}
\email{qhdai@mail.tsinghua.edu.cn}
\authornotemark[1]

\renewcommand\shortauthors{Qiu, Z. et al}

\begin{abstract}
The art of instrument performance stands as a vivid manifestation of human creativity and emotion. Nonetheless, generating instrument performance motions is a highly challenging task, as it requires not only capturing intricate movements but also reconstructing the complex dynamics of the performer-instrument interaction. While existing works primarily focus on modeling partial body motions, we propose Expressive ceLlo performance motion Generation for Audio Rendition (ELGAR), a state-of-the-art diffusion-based framework for whole-body fine-grained instrument performance motion generation solely from audio. To emphasize the interactive nature of the instrument performance, we introduce Hand Interactive Contact Loss (HICL) and Bow Interactive Contact Loss (BICL), which effectively guarantee the authenticity of the interplay. 
Moreover, to better evaluate whether the generated motions align with the semantic context of the music audio, we design novel metrics specifically for string instrument performance motion generation, including finger-contact distance, bow-string distance, and bowing score. Extensive evaluations and ablation studies are conducted to validate the efficacy of the proposed methods. In addition, we put forward a motion generation dataset SPD-GEN, collated and normalized from the MoCap dataset SPD. As demonstrated, ELGAR has shown great potential in generating instrument performance motions with complicated and fast interactions, which will promote further development in areas such as animation, music education, interactive art creation, etc. Our code and SPD-GEN dataset are available at https://github.com/Qzping/ELGAR.
\end{abstract}

%
%


%
%



\begin{teaserfigure}
  \includegraphics[width=\textwidth]{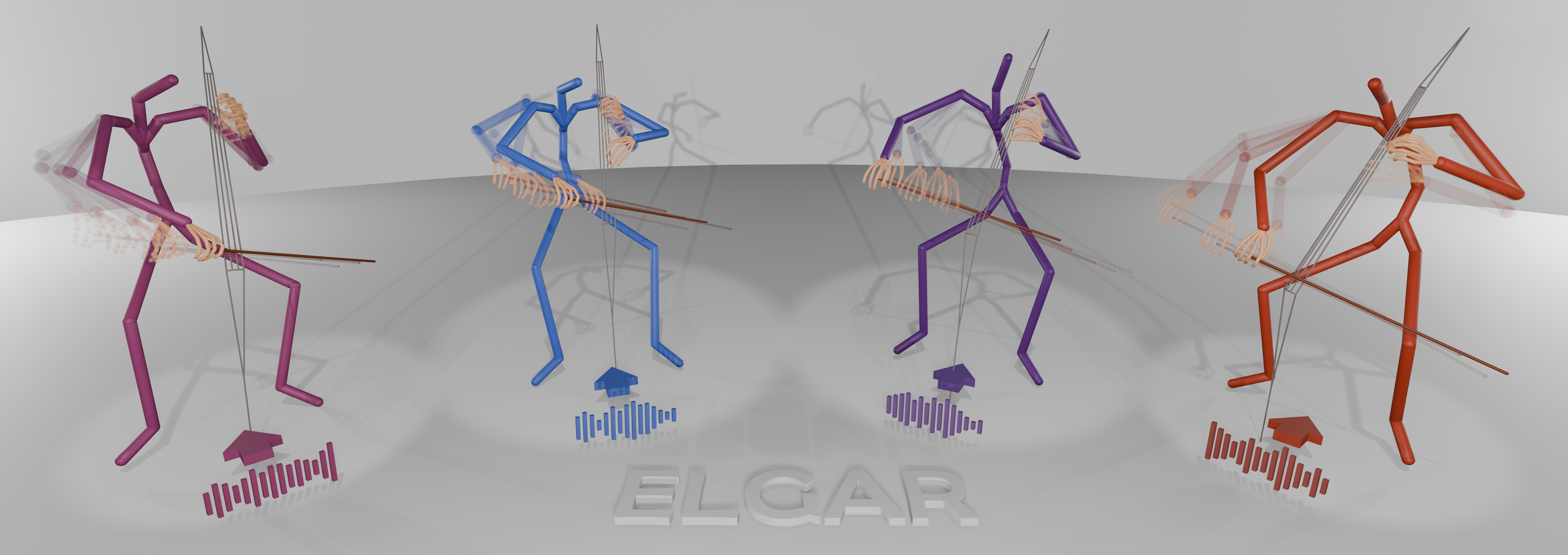}
  \caption{ELGAR is capable of generating cello performance motion with precise details and complicated interactions solely from audio}
  \Description{}
  \label{fig:teaser}
\end{teaserfigure}

\maketitle

\section{Introduction}
Instrument performance, as an art form, carries not only the rich auditory landscape, but also the unspoken language conveyed through the movements of musicians. Every performance encapsulates a dialogue between the performer and their instrument, where subtle gestures and precise motions shape the comprehensive experience. Among the vast array of instruments, string instruments exemplify a delicate interplay of control and expression, featured by continuous (non-discrete) playing positions, in contrast to fixed-interval instruments such as the piano or fretted guitar. In this research, we choose the cello as a representative example for its prominent solo property and a wide pitch range (from bass to soprano) among the violin family. The continuous nature in cello playing requires exquisite coordination between the performer’s hands, the bow, and the strings. Such intricate choreography of motion and interaction is what breathes life into music, yet it still remains one of the most challenging aspects to synthesize a plausible and natural cello-playing motion. In pursuit of this goal, one may take symbolic representations (e.g., sheet music or MIDI) or raw audio as input; either of them represents different dimensions of music rendition. Among these modalities, we focus on generating performance from raw audio, particularly in an end-to-end manner. While audio input is more complex than symbolic music due to its continuous nature and lack of explicit structure, this complexity is more of an asset than a drawback. Audio embeds richer expressive power of performance, as different interpretations of the same musical piece are reflected in each audio recording, making it a particularly valuable modality for performance motion generation. Furthermore, the audio is highly accessible, thanks to abundant online resources and the fact that it requires no specialized musical knowledge to obtain or understand. This combination of expressiveness and accessibility positions audio-based performance generation as a highly promising task with broad application potential.

The rapid advancements in motion generation tasks have brought us closer to realizing this ambitious goal. Leveraging the flourishing breakthroughs in generative AI, such as GANs \cite{karras2019stylegan}, VAEs \cite{razavi2019vqvae2}, Transformers \cite{achiam2023gpt4}, and Diffusion models \cite{peebles2023scalablediff}, a range of works have harnessed diverse cross-modal inputs to generate motion for various scenarios \cite{tevet2022mdm, gong2023tm2d, ng2024audio2photoreal, tseng2023edge}, with certain methods achieving notable progress in fine-grained control \cite{xie2023omnicontrol, karunratanakul2023gmd}, while others push the boundaries of interaction synthesis \cite{li2025controllable, liang2024intergen}.
For performance motion, the generation task becomes even more challenging as it requires not only meeting general motion quality standards but also adhering to musical rules and constraints.

Existing work on generating instrument performance motion can be broadly categorized into two paradigms. The first paradigm employs Supervised Learning \cite{shlizerman2018audio2bodydynamics, kao2020temporally, chen2021guzheng}, relying on pre-collected datasets for training. However, these methods only concentrate on the body motions, failing to account for the nuanced interactions. The second paradigm utilizes Reinforcement Learning (RL) \cite{wang2024furelise, xu2024synchronize} to generate motions that adhere to physical constraints, but it depends on symbolic representations and also relies on a physical simulation environment for training and generation, which limits the applicability to more general scenarios. Furthermore, existing RL-based works are limited to generating partial body motions for performance. As such, end-to-end audio-driven full-body performance motion generation (audio-to-perform) remains untouched, as it involves both precise control over the movements and intricate interaction between the performer and the instrument. Moreover, since instrumental performance is governed by musical regularities, the evaluation of performance motion generation should extend beyond general metrics to account for these performance constraints—an aspect overlooked by existing methods.

In this study, we pioneer a diffusion-based framework for whole-body instrument performance motion generation using audio alone, capable of depicting the fine-grained hand movements and recovering the intricate interactions, dubbed Expressive ceLlo performance motion Generation for Audio Rendition (ELGAR).
To highlight performer-instrument interaction, we introduce Hand Interactive Contact Loss (HICL) and Bow Interactive Contact Loss (BICL), derived from audio cues in the SPD dataset \cite{jin2024audio}. Grounded in the physics of sound production and domain-specific knowledge of the instrument, these tailored losses target the key performance elements while enhancing the accuracy of the spatial relationship between the performer and the instrument.
Existing instrument performance datasets \cite{papiotis2016quartet, volpe2017multimodal, jin2024audio} are not well-suited for motion generation, as they provide only keypoints positions without kinematic information or positional constraints, leading to an overly large solution space. Additionally, variations in body shape further complicate the issue, as there is no unified human body representation for instrument performance motion to generalize across individuals. To address this, we further collate and process the motion capture data from the SPD dataset \cite{jin2024audio}, a high-quality dataset covering performer and instrument motion, resulting in a reliable motion generation dataset SPD-GEN, which can serve as a new benchmark for the task of 3D instrument performance motion generation.

To summarize, our key contributions are: \par
1) To the best of our knowledge, we present the first solution for generating whole-body instrument performance motions featuring fine-grained details and intricate interactions directly from audio signals, marking a novel attempt with promising results for this emerging task. \par
2) We propose Hand Interactive Contact Loss (HICL) and Bow Interactive Contact Loss (BICL), which enhance the realism and plausibility of the generated performance motions. \par
3) We design new metrics for the generation of string performance motion, including finger-contact distance, bow-string distance, and bowing scores. \par
4) We introduce the SPD-GEN dataset, specifically tailored for motion generation tasks.

\section{Related Work}

\subsection{General Motion Generation}

Motion generation has long been an active and continuously evolving research area. Recent advances have been driven by improved techniques for constraining and guiding the generation of movements, often through the integration of rich and nuanced semantic information from multimodal cues. This has enabled the flexible and diverse synthesis of motions that closely align with specific input.
Common modalities include text, speech, music, etc., supporting a wide range of applications such as generating actions from textual descriptions\cite{tevet2022mdm, petrovich2023tmr, kong2023priority, jin2024actasyouwish}, producing natural gestures from speech dynamics\cite{alexanderson2023listen, ng2024audio2photoreal}, and creating dances that match the musical rhythms\cite{alexanderson2023listen, siyao2023bailando++, tseng2023edge}, etc. A considerable number of the mentioned works leverage diffusion models \cite{ho2022classifier-free, ramesh2022hierarchical, peebles2023scalablediff}, whose recent advances have significantly boosted the quality of motion generation.

Building upon multimodal inputs, some methods take motion generation a step further by introducing additional controls, allowing for more delicate and refined motion synthesis\cite{karunratanakul2023gmd, xie2023omnicontrol, cohan2024flexible}. These approaches not only rely on textual inputs to guide motion generation but also integrate spatial constraints, ensuring that the generated motions not only align with the content of the text but also conform to precise spatial signals. Although these works have achieved preliminary controllable generation capabilities, the level of control remains limited, particularly when fine-grained control is required in the context of complex interactive motions.

Moreover, several works extend the capability of motion generation given multimodal prompts by capturing complex and dynamic interactive behaviors, including human-object interaction\cite{cha2024text2hoi, diller2024cg, li2025controllable} and human-human interaction\cite{tanaka2023role, liang2024intergen}. These works offer powerful tools for applications requiring coordination. However, these methods either generate body motions exclusively or focus solely on hand movements, leaving the generation of complex and detailed full-body interactive motions an open problem.


\subsection{Instrument Performance Generation}
Beyond early attempts at instrument performance generation \cite{elkoura2003handrix, zhu2013system}, recent works follow the trend of data-driven methods. \cite{shlizerman2018audio2bodydynamics} investigates the feasibility of generating piano and violin performance motions from audio, asserting that natural body dynamics can be recovered from audio signals. \cite{liu2020bodymovement}, also starting from audio, demonstrates the generation of plausible upper-body violin movements. \cite{li2018skeleton}, on the other hand, utilizes MIDI signal streams to generate piano performance motions online. However, all of these studies focus solely on 2D motions. The first effort to generate 3D violin performance motions was presented by \cite{kao2020temporally}. Using GANs, \cite{chen2021guzheng} showcases the generation of Guzheng performance animations synchronized with input music. The papers mentioned so far rely on Supervised Learning, which results in suboptimal outcomes for generating complex hand movements and overlooks the interaction with the instrument.

Lately, two studies employ Reinforcement Learning (RL) to generate physics-based hand-playing motions for instrument performance\cite{xu2024synchronize, wang2024furelise}. Training is driven by explicit reward functions, enabling complex interactive motions that comply with physical constraints. However, they do not support end-to-end audio-driven generation, relying on symbolic representations as input, which require expertise in music to interpret, and are incapable of producing personalized or stylized motions. Additionally, these RL-based approaches are confined to hand motion generation, and extending them to full-body motion would likely require coordinating more agents, thereby significantly increasing the complexity of the task.

Previous works on instrument performance generation have focused on partial body motions, typically limited to the torso or hands. In contrast, we introduce whole-body performance motion generation, encompassing intricate hand movements and bowing action for a more comprehensive modeling.

\begin{figure}[ht]
    \centering
    \includegraphics[width=\the\columnwidth]{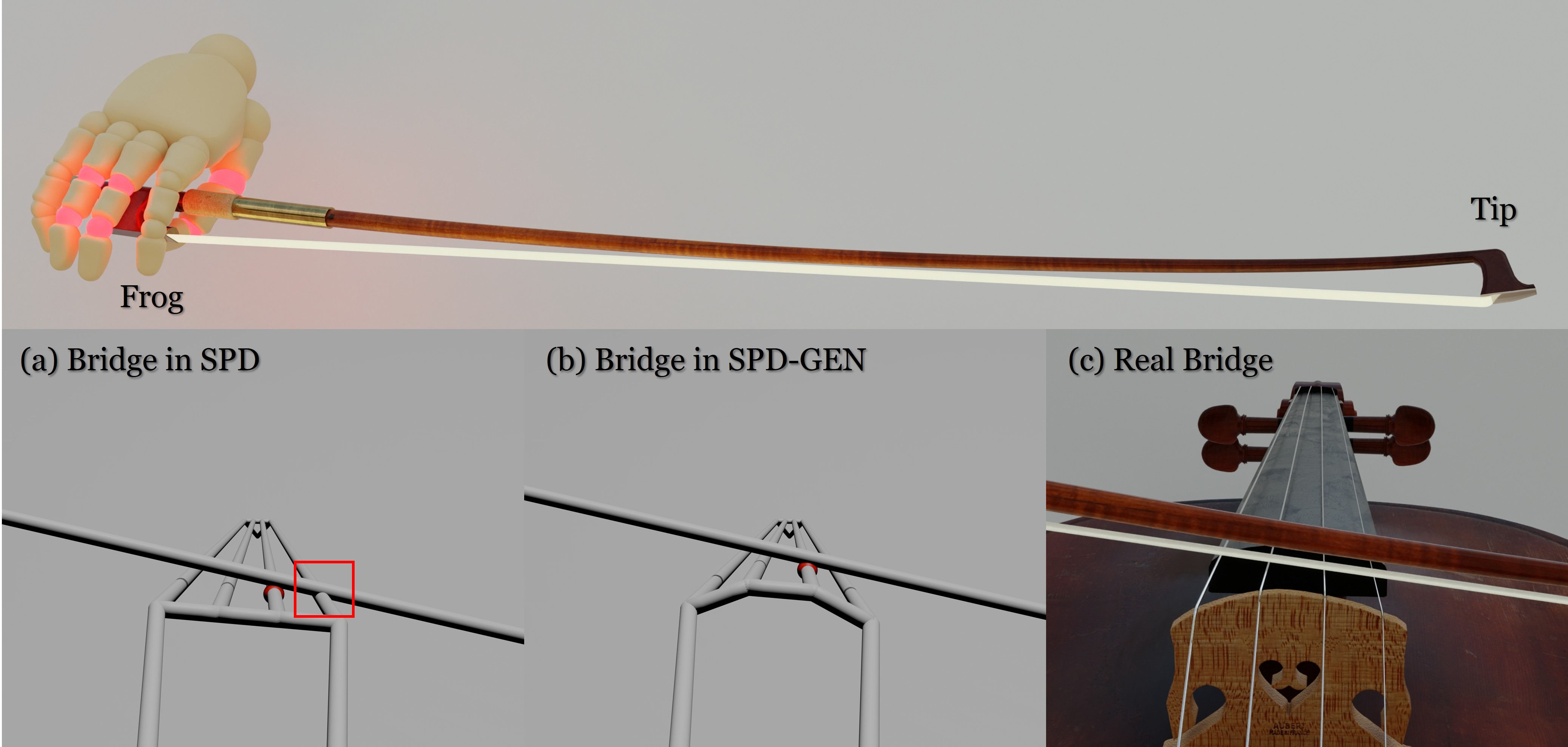}
    \caption
        {
            \textbf{Top:} We position the starting point of the bow (frog) at the midpoint between the PIP and DIP joints of the middle finger, ring finger, and thumb (highlighted in red).
            \textbf{Bottom:} As shown in (b), SPD-GEN reconstructs the arched cello bridge, unlike the flat bridge in SPD, closely matching the actual instrument illustrated in (c). This enables the performer to play the two middle strings without unintended contact with adjacent ones, thereby avoiding potential penetration artifacts as seen in (a). The red dot in (a) and (b) indicates the bow-string contact point.
        }
    \label{illustration}
\end{figure}

\begin{figure*}[t]
    \centering
    \includegraphics[width=\textwidth]{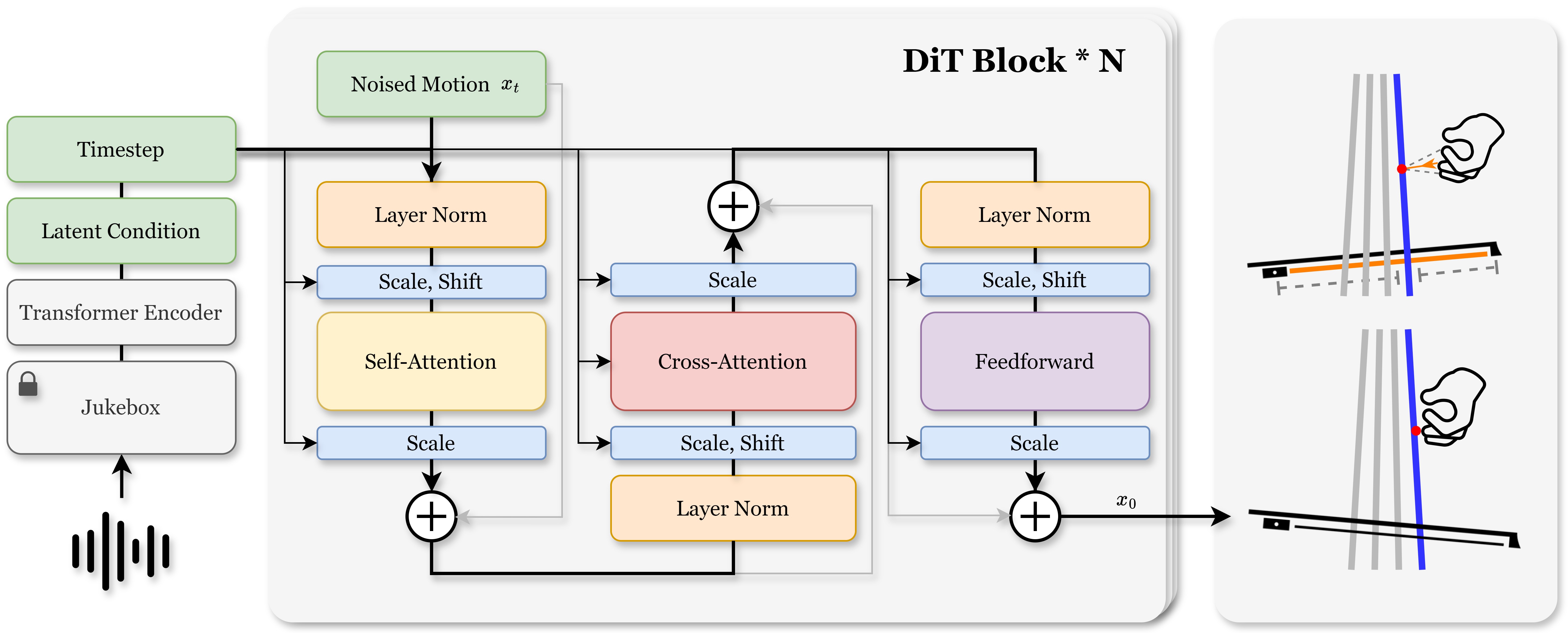}
    \caption
        {
            Given performance audio, ELGAR employs DiT blocks with adaLN-Zero to denoise the performance motions from $x_t$ to $x_0$, incorporating cross-attention to further integrate audio features extracted by a frozen Jukebox \cite{dhariwal2020jukebox}. The upper-right details the \textit{Interactive Contact Loss (ICL)}.
            The orange solid lines represent the "contact" of \textit{ICL}; while the gray dashed lines show the "interactive" relationship between the non-playing fingers and the contact point, as well as between the bow endpoints and the activating string. The Red dot marks the contact position of the hand, and the blue-highlighted string denotes the activating string. For the hand, the note-playing finger should strive to contact the audio-designated position, while other fingers are expected to maintain proper spatial relationships with the contact position. Similarly, the bow must contact the activating string while maintaining proper distance relationships between its two ends and the string.
            \label{architecture}
        }
    \Description{
        <long description>
    }
\end{figure*}

\section{Methodology}

\subsection{Data Preprocess}
The SPD dataset \cite{jin2024audio} contains 81 cello performance pieces by performers of varying height and gender, and the instruments used also differ in shape and placement. To ensure consistency in motion generation, we need to normalize the data as if all pieces were performed by the same person on the same cello.

For cello normalization, we selected a manually labeled cello as the shared instrument in the SPD-GEN dataset. We also restored the arched cello bridge, shown in Figure \ref{illustration}, to better match that of a real cello. This adjustment allows the generated motions to theoretically play the two middle strings without artifacts, which would otherwise occur with a flat bridge. The cellos are then aligned at the end pin position across all frames. We apply the Kabsch algorithm \cite{kabsch1976solution} to efficiently compute the optimal rotation matrix for aligning each frame's cello with the shared cello, while simultaneously transforming all whole-body keypoints.

For human normalization, we employ a two-stage inverse kinematics (IK) process using VPoser in SMPL-X format \cite{SMPL-X:2019}. In the first stage, we perform an initial IK on the human body to fit the average body shape, global orientation, and translation across all frames. In the second stage, we refine the IK using the average body from the previous stage, prioritizing accurate wrist fitting while leaving the elbow and shoulder keypoints unfitted. This strategy allows us to leverage the global wrist rotation provided in the SPD dataset and keep the local rotations of the hand joints.

After the aforementioned data processing, we obtain our SPD-GEN, totaling about 7000 seconds
of whole-body cello performance motion data represented in 6D rotations \cite{zhou2019sixD}. The body comprises 21 joints, excluding the pelvis, while each hand includes 15 joints, yielding $r \in \mathbb{R}^{306}$, where $306=(21+15+15) \cdot 6$. The bow direction is represented by a unit vector $\hat{\mathbf{v}} \in \mathbb{R}^3$. As illustrated in Figure \ref{illustration}, the starting point of the bow, also known as the frog, is anchored between the middle finger, ring finger, and thumb of the left hand. Thus, its endpoint, referred to as the tip, can be determined from the frog and the unit direction vector, given the fixed bow length. As a result, the complete motion representation is $x=\{r,\hat{\mathbf{v}}\} \in \mathbb{R}^{309}$, where $309=306+3$.

\begin{figure*}[t]
    \centering
    \includegraphics[width=\textwidth]{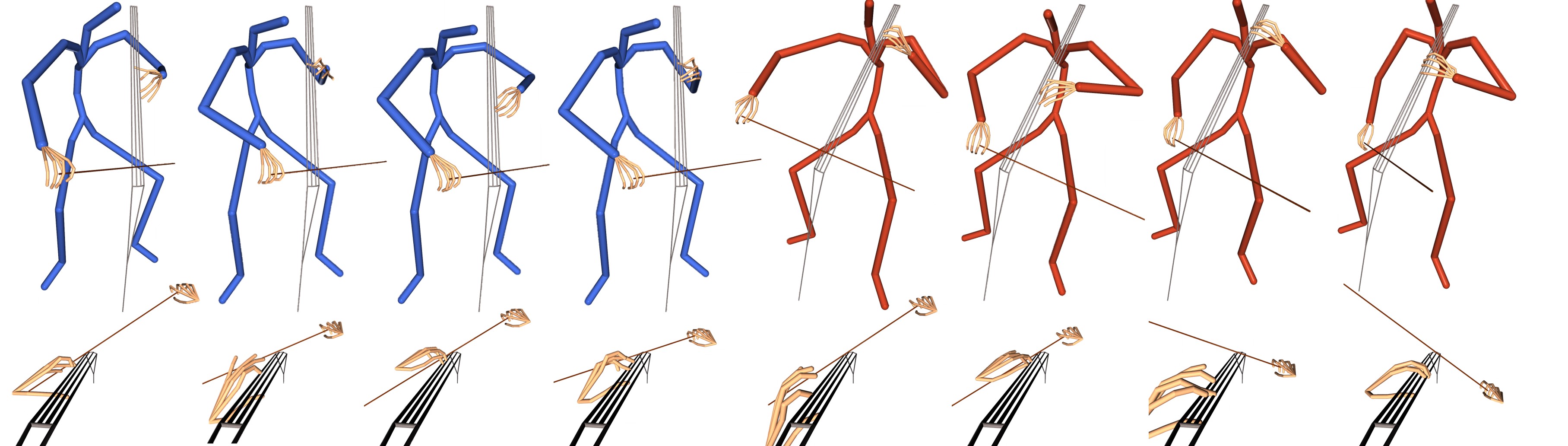}
    \caption
        {
            A variety of sample motions generated by ELGAR, shown from the holistic view and the performer's view, to reveal their diversity and richness.
        }
    \label{demofig}
\end{figure*}

\subsection{Diffusion Preliminaries} \label{sec:diff}
Given the collected dataset, the diffusion model follows the Markov chain, gradually adding random noise to the sample of cello performance motion $x_0 \sim q(x)$, also known as the forward process. By applying the reparameterization trick, we can formulate the process as $x_t$ sample from $x_0$:
\begin{equation}
    q(x_t|x_0)=\sqrt{\bar{\alpha_t}}x_0 + \epsilon\sqrt{1-\bar{\alpha_t}}, \epsilon\sim\mathcal{N}(0,1)
\end{equation}
where $\alpha_t=1-\beta_t$ and $\bar{\alpha_t}=\prod^t_{s=1}\alpha_s$. Constants $\beta_{1:T}$ are hyperparameters. Then, to invert the forward process, the diffusion model learns the backward process to remove the noise from $x_t$:
\begin{equation}
    p_{\theta}(x_{t-1}|x_t) = \mathcal{N}(\mu_{\theta}(x_t), \Sigma_{\theta}(x_t))
\end{equation}
where $\theta$ denotes the model parameters in the neural network, i.e., the transformer in our framework shown in Figure \ref{architecture}.

To condition the generation on musical audio, we explore two potential approaches: \textit{Classifier Guidance} (\textit{CG}) \cite{dhariwal2021classifier-guidance} and \textit{Classifier-Free Guidance} (\textit{CFG}) \cite{ho2022classifier-free}. \textit{CG} facilitates the integration of conditions at inference time, delivering notable results in motion generation with spatial constraints \cite{xie2023omnicontrol, zhang2024rohm}. However, audio condition does not possess the same level of explicit constraints as the spatial condition, making it challenging to develop a function that can effectively approximate a classifier. In addition, \textit{CFG} has demonstrated superior performance over inference-time techniques \cite{nichol2021glide, ramesh2022hierarchical}. Hence, we follow the trend of \textit{CFG}, incorporating the audio condition $c$ during training. In line with \cite{ramesh2022hierarchical}, we train our model to directly predict the motion using a simple mean-squared error loss:
\begin{equation}
    \mathcal{L}_{simple}=\mathbb{E}_{t \sim [1,T], x_t\sim q}[\Vert f_{\theta}(x_t, t, c)-x_0 \Vert]
\end{equation}
Additionally, \textit{CFG} proposes to train an unconditioned model simultaneously by randomly setting the condition $c=\emptyset$ (10\% in our case). On top of that, the sampling procedure applies the following linear combination of the conditional and unconditional generated motions by $w$:
\begin{equation}
    f_{\theta}(x_t, t, c) = (1+w)f_{\theta}(x_t, t, c) - wf_{\theta}(x_t, t, \emptyset)
\end{equation}

\begin{figure*}[t]
    \centering
    \includegraphics[width=\textwidth]{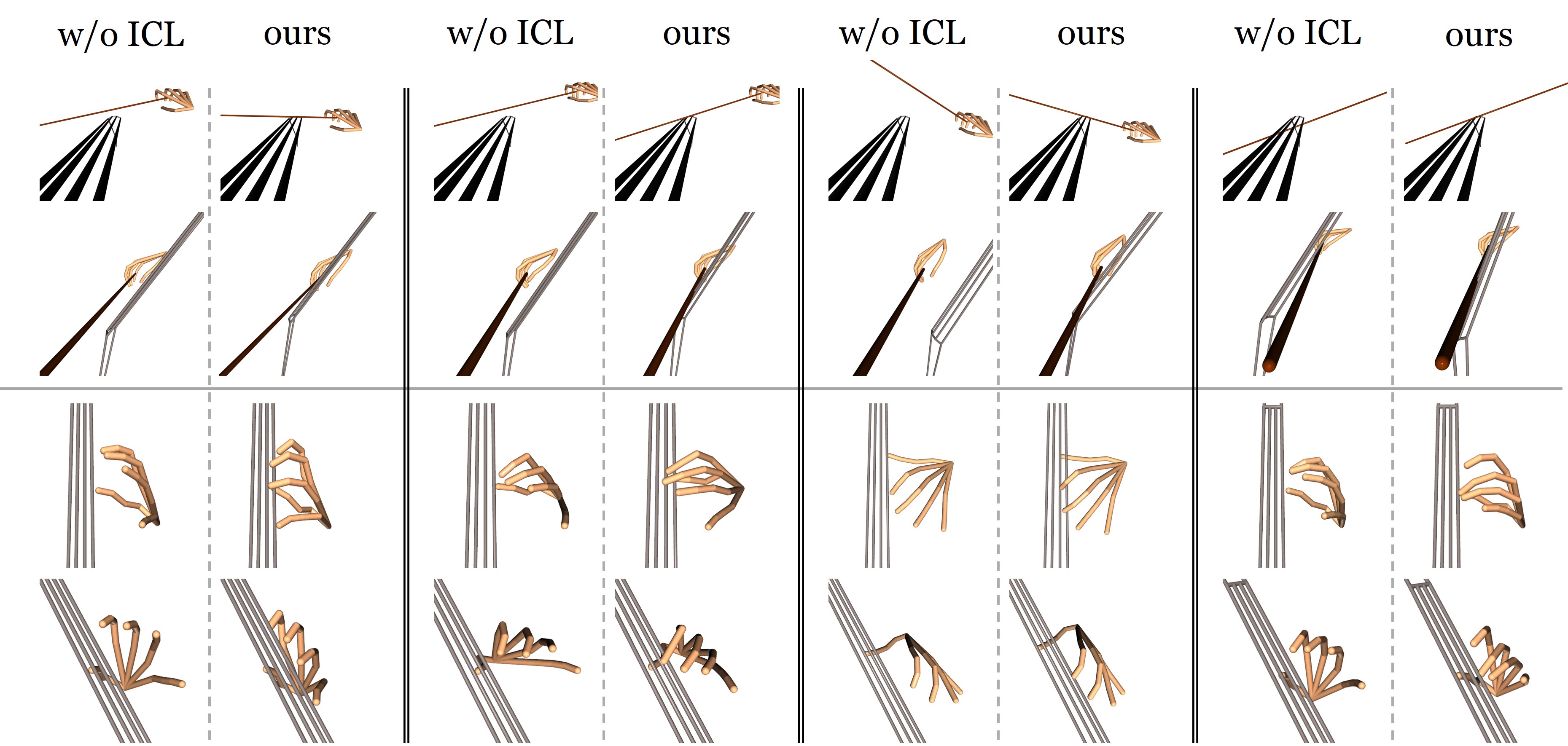}
    \caption
        {
            In this figure, we present a comparative demonstration of the bow and left hand motions before and after the introduction of the \textit{Interactive Contact Loss (ICL)}, highlighting its significant impact. Prior to adopting \textit{ICL}, both the bow and the left hand exhibited noticeable and unrealistic positional deviations relative to the strings. Following the integration of \textit{ICL}, the bow and the left hand display more accurate and reasonable interactions with the strings, aligning closely with the intended playing positions.
        }
    \label{compare}
\end{figure*}

\subsection{Realism Losses}

In our work, we further regularize the generative model by incorporating losses designed to ensure the realism of generated motions, comprising \textit{Geometric Losses} and \textit{Interactive Contact Losses}.

\subsubsection{Geometric Loss}
In the field of motion generation, geometric losses are commonly added to provide the physical regularization \cite{petrovich2021action, tevet2022mdm}. We incorporate four of them to impose constraints on the physical plausibility: 1) position loss, 2) foot contact loss, 3) rotation velocity loss, and 4) position velocity loss. The formulations of position loss, foot contact loss, and rotation velocity loss remain the same as those in \cite{tevet2022mdm}. We further address time coherence on human keypoints and bow keypoints by applying a velocity loss (Eq. (\ref{eq:posvel})) to their positions, attained by the forward kinematic function denoted by $FK(\cdot)$.

\begin{equation} \label{eq:posvel}
    \mathcal{L}_{posvel} = \frac{1}{N-1} \sum_{i=1}^{N-1} \big\| \Delta FK(x_0)^i - \Delta FK(\hat{x_0})^i \big\|_2^2
\end{equation}
$\Delta FK(x_0)^i$ and $\Delta FK(\hat{x_0})^i$ represent the frame-wise positional differences of ground truth motions and generated motions, respectively.

\subsubsection{Interactive Contact Loss}
To better align the generated performance motion with the actual playing, we introduce \textit{Interactive Contact Loss (ICL)}, drawing from domain-specific knowledge of cello performance, which encompasses \textit{Hand Interactive Contact Loss (HICL)} and \textit{Bow Interactive Contact Loss (BICL)}

In cello performance, the left hand plays a crucial role in adjusting the intended pitch. By pressing the string with the note-playing finger, the vibrating length of the string that actually produces the sound varies during the performance. Thus, the fingers of the left hand, particularly the note-playing finger, are essential for cello performance and must conform to specific rules and patterns: 1) the note-playing finger should hold contact with the string, except in the case of open-string playing, and 2) the rest of the fingers should avoid contact with the position pressed by note-playing finger while preserving a natural playing gesture. To address these constraints, we present \textit{HICL} by drawing inspiration from foot contact loss from \textit{Geometric Losses} and the distance map loss introduced in \cite{liang2024intergen}. \textit{HICL} leverages theoretical contact position on strings extracted from audio \cite{jin2024audio}, enforcing restrictions on the contact of the note-playing finger with the string and the interactive relationships of the non-playing fingertips relative to the string.

Accordingly, we acquire our \textit{HICL}, as illustrated below:
\begin{equation}
    \mathcal{L}_{hand} = \mathbb{1}_{note} \Vert \hat{d}_{cp} \odot I_{f_0} \Vert_2^2 
+ \mathbb{1}_{others} \Vert (\hat{d}_{cp} - d_{cp}) \odot I_{f_0} \Vert_2^2
\end{equation}
where $\mathbb{1}_{note}$ and $\mathbb{1}_{others}$ indicate whether the finger is the note-playing finger. $ \hat{d}_{cp}$ represents the predicted fingertip-to-contact distance, while $d_{cp}$ denotes the ground truth distance. $I(\cdot)$ is also an indicator function that activates the loss when a pitch, namely the fundamental frequency $f_0$, is detected.

The bow, held by the right hand, is also an indispensable part of cello performance. Given the pitch determined by the left hand, the right hand maneuvers the bow to excite the strings, thereby shaping the overall performance. Hence, the bow must maintain "contact" with the string to induce vibration. In addition, the bow "interacts" with the string by moving back and forth perpendicularly across the string, guided by both the musical phrase itself and the performer's interpretation.

Although the bow placement is not as explicit as the hand placement given audio, the activating string can still serve as a significant constraint to embody the bowing characteristics mentioned above in the generated motion. Thus, we employ \textit{BICL}, as shown below:
\begin{equation}
    \mathcal{L}_{bow} = \Vert \hat{d_{l_s,l_b}} \odot I_{f_0} \Vert_2^2 
+ \Vert (\hat{d_{p,l_s}} - d_{p,l_s}) \odot I_{f_0} \Vert_2^2
\end{equation}
where $\hat{d_{l_s,l_b}}$ is defined as the distance between the activating string and the predicted bow. $\hat{d_{p,l_s}}$ and $d_{p,l_s}$ are the distances between the bow endpoints and the playing string. They are also controlled by the indicator function $I_{f_0}$.

We find that both \textit{HICL} and \textit{BICL} considerably refine the generated motions, as demonstrated in Section \ref{evaluations}.
In summary, the overall loss is formulated as follows:
\begin{equation}
    \begin{aligned}
        \mathcal{L} = \lambda_{simple}\mathcal{L}_{simple} 
        + \lambda_{foot}\mathcal{L}_{foot}
        + \lambda_{pos}\mathcal{L}_{pos}  
        + \lambda_{rotvel}\mathcal{L}_{rotvel} \\
        + \lambda_{posvel}\mathcal{L}_{posvel}  
        + \lambda_{hand}\mathcal{L}_{hand} 
        + \lambda_{bow}\mathcal{L}_{bow}
    \end{aligned}
\end{equation}

\subsection{Framework}

Our framework is outlined in Figure \ref{architecture}. We first leverage a frozen Jukebox model \cite{dhariwal2020jukebox} as the encoder for the audio condition, as its extracted audio features have been proven robust in various tasks \cite{castellon2021codified, wei2024domusic, tseng2023edge}. Subsequently, a denoising network is required, the $f_{\theta}$ in Section \ref{sec:diff}, given the encoded representation, the performance motion with noise, and the timestep information. Inspired by \cite{tseng2023edge, saharia2022photorealistic}, our denoising structure builds upon the Transformer Decoder to enhance the integration of extracted audio features into the denoising process through cross-attention mechanisms. In addition, we refer to DiT with the adaLN-Zero block \cite{peebles2023scalablediff}, which has shown exceptional capability in class-conditional image generation. Compared to the adaLN block (e.g., the FiLM block in EDGE \cite{tseng2023edge}), the adaLN-Zero block regresses dimension-wise scaling parameters that are applied immediately prior to any residual connections within the DiT block.

\section{Experiment}

\subsection{Implementation Details}
We implement our model with 8 DiT blocks, totaling 55M parameters with a latent dimension of 512. We train our diffusion with 1000 timesteps, and DDIM \cite{song2020ddim} sampling is applied to accelerate the generation with 50 timesteps. Rather than the linear schedule, we add noise by the cosine schedule \cite{nichol2021improved}. Given the limited training data, we train our model with a batch size of 64 on an NVIDIA H800 GPU for 90,000 steps.

We slice our data into 5-second segments for training. For segments shorter than five seconds, the motion of the final frame is used as padding. In order to generate a longer sequence of performance motion, we follow the long-form sampling strategy \cite{tseng2023edge} by leveraging the train-free editability commonly used in motion in-betweening tasks. To further enhance the consistency of the performance motions across different slices, we overlap 4 seconds between the two slices and perform a linearly decaying weighted sum.

\subsection{Evaluations} \label{evaluations}

Figure \ref{demofig} comprehensively demonstrates how ELGAR performs in cello performance generation from multiple perspectives, showing it in a reasonable, accurate, and vivid manner. Evaluation from both qualitative and quantitative perspectives is conducted, by comparing and analyzing the generation results across various training configurations, thereby demonstrating the necessity of each component. In Figure 5, we present a visual comparison of the motion and the interaction with the instrument, using consistent audio input with temporally aligned frames.

Previous works focusing on motion generation often utilize the Fréchet Inception Distance (FID) as a metric \cite{guo2022generating, tevet2022mdm} to assess the overall quality, measuring the discrepancy between the distribution of the generated motions and that of ground truth motions. However, we argue that FID is not well-suited for our task. First, we incorporate additional constraints (HICL and BICL) during training, integrating audio information beyond the motions. This naturally leads to distributional differences between the generated motions and the dataset motions. A similar case has been shown in the prior work \cite{tevet2022mdm}, where the introduction of foot contact loss yielded visually better but metrically worse results. Second, the SPD-GEN dataset is relatively small in size, sharing an issue in AIST++ noted by \cite{tseng2023edge}, where the test set fails to fully represent the motion distribution of the training set.

Consequently, to more specifically evaluate our results targeting the key elements of string performance motion, we introduce several novel metrics: finger-contact distance, bow-string distance, and bowing scores, which are grounded in the domain knowledge of string instrument performance. 

The first two metrics are designed to evaluate whether our results accurately replicate the physical interactions between the performer and the instrument.
The finger-contact distance examines the deviation between the tip of the left-hand note-playing finger and the trigger position on the cello for the current pitch. 
While there are various reasonable performance motions for a given segment of cello music, as most notes can be played using different techniques across different strings, we determine the trigger position closest to the performer’s note-playing finger as the "intent" of the generated motion. 
Shorter finger-contact distance indicates a more accurate and appropriate performance motion.
The bow-string distance reflects the deviation between the bow and the string to be struck, which is uniquely identified once the aforementioned trigger position is determined.
A smaller deviation indicates a more accurate reproduction of the interplay in which the bow excites the string’s vibration.

Although there is no strict rule for bow change timing, certain moments are musically more appropriate in terms of rhythm and phrasing. The bowing F1-score examines whether the generated motion aligns with these musically suitable bowing attacks, reflecting the model’s ability to detect audio-driven bowing cues. We use 10\% of SPD-GEN as the test set and detect bowing attacks in both ground-truth and the generated motions by analyzing the movement direction of the bow frog relative to the bridge. Following \cite{kao2020temporally}, a tolerance $\delta$ of 3 frames (0.1 seconds) is applied: if a predicted attack falls within $[i - \delta, i + \delta ]$ of a ground-truth bowing attack $A(i)$, it counts as a true positive. On the other hand, to further assess to what extent the bowing patterns align with human performance, we compute the cosine similarity of the relative distance between the bow and the string across temporal dimension, where the relative position is negative when the lower half of the bow strikes the string and positive when the upper half does.

In the ablation study, the introduction of the HICL and the BICL significantly improves the performance of both the fingering hand and bowing hand, demonstrating their validity in the string performance generation task. 
The evaluation results based on the aforementioned metrics are shown in Table \ref{tab:ablation}, with the corresponding visual comparisons illustrated in Figure \ref{compare} and the supplementary video.

\begin{table}[t]
    \caption{Ablation study showing the impact of including or excluding HICL and BICL on the generated results across the metrics of Finger-Contact Distance (FCD, in mm), Bow-String Distance (BSD, in mm), Bowing F1-Score (BF1), and Bowing Cosine Similarity (BCS). 
     \textbf{Bold} indicates best result.
    }
    
    \centering
    \resizebox{\columnwidth}{!}
    {
    \begin{tabular}{l|llll}
    \toprule
    Loss Configuration  & FCD $\downarrow$ & BSD $\downarrow$ & BF1 $\uparrow$ & BCS $\uparrow$  \\
    \midrule
    w/o ICL & 18.64  & 25.20  & 0.4332 & 0.6965  \\
    w/ HICL only  & \textbf{14.56} & 23.98 & 0.4082 & 0.6646  \\
    w/ both HICL and BICL  & 15.60 & \textbf{5.40} & \textbf{0.4721} & \textbf{0.7515} \\
    \bottomrule
    \end{tabular}
    }
    
    \label{tab:ablation}
\end{table}

\section{Discussion}
In real-world instrument performance, current playing motions are renditions of both past and future music context, rather than instantaneous decisions. While ELGAR maintains general musical coherence and follows adequate performance conventions, its limited context awareness in long sequences can still lead to unnatural bow transitions during sustained passages. Indeed, a better approach is expected for generating long performance motions. By incorporating more contextual cues, the generated long-sequence performance motions could be more coherent and plausible.

Additionally, for our contact-related task, directly predicting joint location could be a competitive alternative to our current joint rotation choice, as it provides a more straightforward way to enforce contact constraints. Even so, generating rotations facilitates easier integration with animation. A better combination of these two representations is worth discovering in the future.

Admittedly, our model simplifies finger-string pressure into binary states (i.e., pressed and unpressed), yet it sufficiently covers most playing scenarios. While modeling finer pressure could enhance realism, it requires additional modalities (e.g., force sensors) beyond pose or position data, which are currently unavailable. 
Another limitation lies in the static cello assumption, while real performance involves natural instrument dynamics, as illustrated by artist-adjusted renderings in the supplementary video. It is also worth noting that, even with the BICL constraint, the bow occasionally loses contact with the strings, suggesting the need for more robust strategies. We leave these as future work.

Furthermore, while the SPD-GEN dataset already encompasses a rich variety of cello performance motions, it remains limited in scale, particularly in terms of diverse performance styles for the same musical piece. The absence of such data restricts our work to a narrower range of stylized expressiveness and hinders the ability to capture the full spectrum of possible performance variations.

Moreover, applying generated instrument performance motions to downstream scenarios such as films and games poses a critical challenge of retargeting the SMPL-X motions to various virtual characters. While commercial animation tools like Unreal Engine provide solutions for retargeting, they often neglect the interactive aspects, resulting in artifacts. Animators have to meticulously craft the poses and gestures of characters to mitigate this effect. Recent academic papers have proposed several methods for interactive motion retargeting \cite{jin2018aura, zhang2023simulation, jang2024geometry}. Yet apparently, these approaches are not applicable to motions featuring rich, sophisticated, and fine-grained interactions. Our work, by generating such motions that contain precise interactions, offers a novel and challenging scenario for the retargeting research field. If the interaction details between the hand and the instrument can be accurately preserved after retargeting, it would significantly reduce the cost of animation production, which could be further adapted to other human-object interaction scenarios.

\section{Conclusion}
To conclude, we propose ELGAR, a diffusion-based approach for cello performance motion generation solely from audio input. To the best of our knowledge, this is the first study to achieve whole-body motion synthesis for musical instrument performance, excelling in generating fine-grained motions and reconstructing intricate interactions. We further present the Hand Interactive Contact Loss (HICL) and Bow Interactive Contact Loss (BICL), maintaining the fidelity of the interplay between the performer and the instrument. Additionally, dedicated metrics for string performance are introduced to better evaluate the generated motions, including finger-contact distance, bow-string distance, and bowing scores. On top of these, we contribute SPD-GEN, a motion generation dataset derived from the motion capture dataset SPD. Through experiments, ELGAR has been proven to generate high-quality, realistic performance motions with complex and dynamic interactions.
As illustrated, ELGAR opens up multiple promising pathways for future work, offering novel insights and inspiration for the research field and advancing a wide spectrum of applications.

\begin{acks}
This work was supported in part by the National Key R\&D Program of China (No.2024YFB2809101), in part by the NSFC (No.62171255), in part by the Tsinghua University - Migu Xinkong Culture Technology (Xiamen) Co.Ltd. Joint Research Center for Intelligent Light Field and Interaction Technology, PhaseII, in part by the Guoqiang Institute of Tsinghua University (No.2021GQG0001), in part by the Special Program of National Natural Science Foundation of China (Grant No. T2341003), in part by the Advanced Discipline Construction Project of Beijing Universities, in part by the Major Program of National Social Science Fund of China (Grant No. 21ZD19), in part by the Key Research Program of Central Conservatory of Music (NO.24ZD04).
\end{acks}

\bibliographystyle{ACM-Reference-Format}
\bibliography{2_Reference}

\end{document}